\documentclass[preprint,12pt]{elsarticle}

\usepackage{amsmath}
\usepackage{amsfonts}
\usepackage{amssymb}
\usepackage{calligra}
\usepackage{calrsfs}
\usepackage{color}
\usepackage{tcolorbox}
\usepackage{subcaption}
\usepackage{titlesec}
\usepackage{geometry}
\usepackage[ruled]{algorithm2e}

\journal{Mechanics of Materials}

\DeclareMathOperator*{\argmin}{arg\,min}

\newtheorem{theorem}{Theorem}

\begin{document}

\begin{frontmatter}

\title{Imposing equilibrium  on  measured  3-D  stress  fields  using  Hodge  decomposition  and  FFT-based optimization}

\author[1]{Hao Zhou}
\author[2]{Ricardo A. Lebensohn}
\author[3,4]{P\'eter Reischig}
\author[3,5]{Wolfgang Ludwig}
\author[1]{Kaushik Bhattacharya}
\affiliation[1]{Division of Engineering and Applied Science, California Institute of Technology, Pasadena CA 91125, USA}
\affiliation[2]{Theoretical Division, Los Alamos National Laboratory, Los Alamos, NM 87544, USA}
\affiliation[3]{European Synchrotron Radiation Facility, Grenoble, France}
\affiliation[4]{InnoCryst Ltd, Daresbury, Warrington, WA4 4AD, UK}
\affiliation[5]{Laboratoire Materiaux, Ingenierie et Science, INSA Lyon, Lyon, France}

\begin{abstract}
We present a methodology to impose micromechanical constraints, i.e. stress equilibrium at grain and sub-grain scale, to an arbitrary (non-equilibrated) voxelized stress field obtained, for example, by means of synchrotron X-ray diffraction techniques. The method consists in finding the equilibrated stress field closest (in $L^2$-norm sense) to the measured non-equilibrated stress field, via the solution of an optimization problem. The extraction of the divergence-free (equilibrated) part of a general (non-equilibrated) field is performed using the Hodge decomposition of a symmetric matrix field, which is the generalization of the Helmholtz decomposition of a vector field into the sum of an irrotational field and a solenoidal field. The combination of: a) the Euler-Lagrange equations that solve the optimization problem, and b) the Hodge decomposition, gives a differential expression that contains the bi-harmonic operator and two times the curl operator acting on the measured stress field. These high-order derivatives can be efficiently performed in Fourier space. The method is applied to filter the non-equilibrated parts of a synthetic piecewise constant stress fields with a known ground truth, and stress fields in Gum Metal, a beta-Ti-based alloy measured in-situ using Diffraction Contrast Tomography (DCT). In both cases, the largest corrections were obtained near grain boundaries.
\end{abstract}

%%Graphical abstract
\begin{graphicalabstract}
   	\begin{center}
   		\resizebox{.99\textwidth}{!}{%
   			\includegraphics[height=3cm]{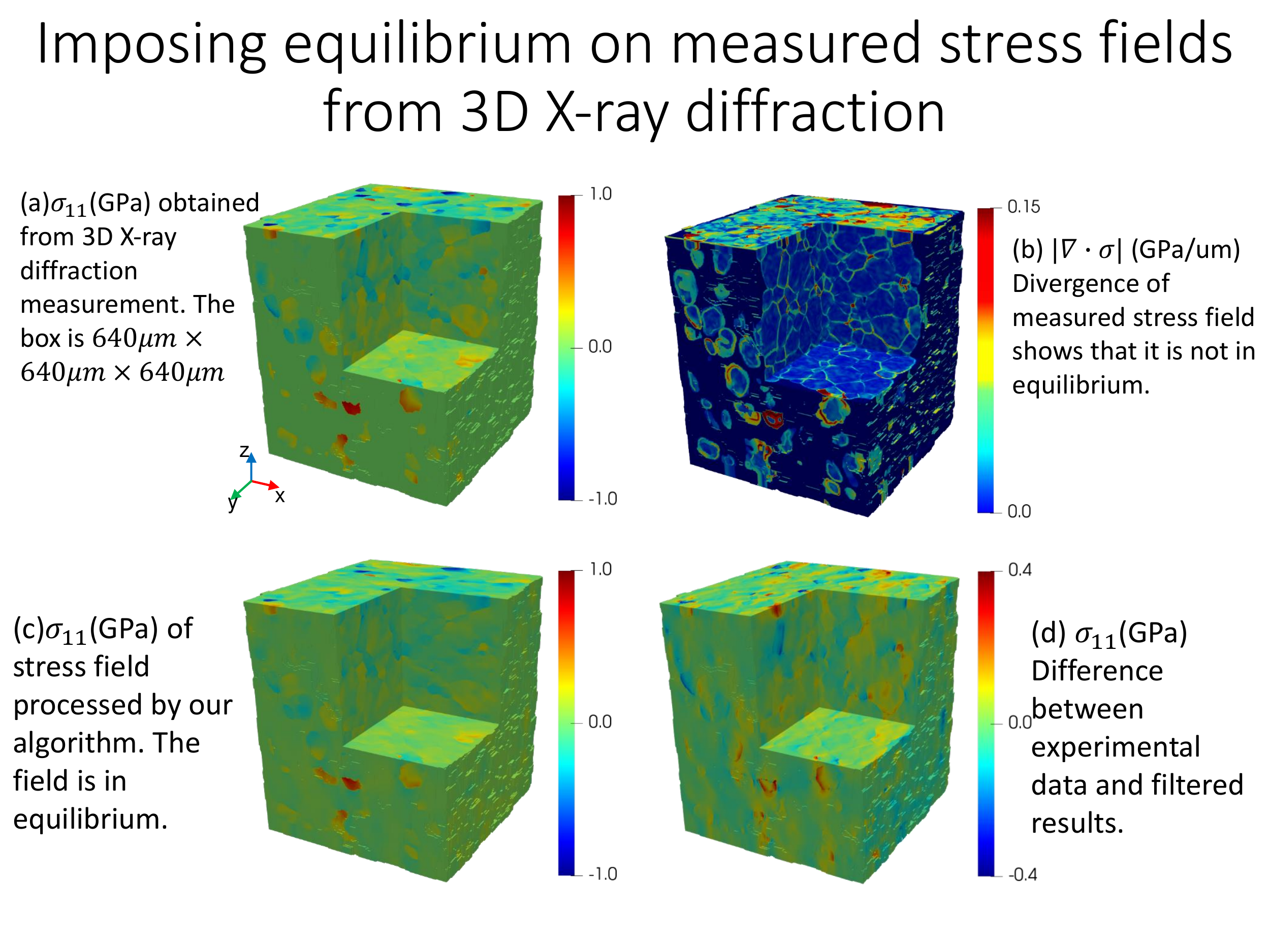}
   		}
   	\end{center}
\end{graphicalabstract}

%%Research highlights
\begin{highlights}
\item Analytical formulation to impose stress equilibrium is obtained through a variational optimization method.
\item A FFT-based numerical solver is developed and further modified for interface discontinuities.
\item The algorithm is first applied to synthetic stress fields, which provide a known ground truth.
\item The algorithm is applied to a 3-D X-ray diffraction experiment of a beta-Ti alloy, resulting in large corrections near grain boundaries.
\end{highlights}

\begin{keyword}
%% keywords here, in the form: keyword \sep keyword
X-ray diffraction \sep experimental mechanics \sep stress equilibrium \sep Hodge decomposition \sep fast Fourier transform \sep micromechanics
%{\color{blue} could not find a right tag in PACS for our paper}
%% PACS codes here, in the form: \PACS code \sep code
%\PACS 0000 \sep 1111
%% MSC codes here, in the form: \MSC code \sep code
%% or \MSC[2008] code \sep code (2000 is the default)
%\MSC 0000 \sep 1111
\end{keyword}

\end{frontmatter}

\section{Introduction}

%To recover the stress field, a straightforward way is through elastic constitutive relations.
%For example, residual stress comes into mechanical parts in the process of manufacturing. Reasons includes temperature change, inelastic deformation and phase change. Residual stress can significantly enhance or deteriorate the performance of materials. Thus it becomes of importance to accurately characterize the stress distribution.

Emerging non-destructive X-ray diffraction (XRD) techniques for 3-D material characterization performed at light sources, collectively known as 3DXRD (e.g. \cite{poulsen2004three,ludwig2009three,lienert2011high}), provide previously inaccessible in-situ microstructural and micromechanical information on polycrystalline materials at grain and sub-grain scales. In-situ 3-D measurements are enabled by the collection of multiple diffraction patterns as the sample is rotated and deformed. These patterns are then used in combination with forward-modelling of the diffraction experiment to optimize the agreement between those measured detector images and simulated patterns corresponding to every possible orientation and micromechanical field configuration. Near-field (nf) techniques \cite{suter2006forward,li2013adaptive,vigano2016three} are used to obtain crystal orientation fields, resulting in space-resolved voxelized microstructural images with intragranular resolution. Further, far-field (ff) techniques \cite{poulsen2001three,bernier2011far,oddershede2010determining} provide information on local micromechanical fields. Due to a trade-off between direct and reciprocal space resolutions, the original ff data inversion methods were able to deliver average stresses/elastic strains in single crystal grains and the volume and location of the centers of mass of those grains, but not voxelized intragranular fields.

Recently, 3DXRD data inversion methods were improved to provide not only voxelized crystal orientation but also stress fields with intragranular resolution using a variety of experimental procedures, including a microbeam (1D), line beam (2D) or box beam (3D) illumination, each demanding a tailored optimization-based data processing approach \cite{reischig2019three,Hayashi2019, reischig2020three,shen2020voxel,Henningsson2020}. However, these types of adopted optimization methods have been exclusively based on imposing diffraction constraints, by minimizing the difference between experimental and simulated detector images obtained by forward-modelling of the diffraction experiment, and retaining as {\it measured} fields the orientation and stress fields that produced that minimum. In either case, using the original data inversion based on grain averages, or the more advanced methods that deliver voxelized information, the resulting stress fields are in general unbalanced (not divergence-free, violating the stress equilibrium condition).

%RLThe equilibrium of moments acting on a small local region, such as a voxel, is inherently enforced by the symmetry of the stress tensor. 
Since the micro-mechanical field that directly affects diffraction patterns---by changing the local lattice spacing---is elastic strain, accounting for stress equilibrium at the sub-grain level requires a fairly accurate knowledge of the single crystal properties, such as zero-stress lattice parameters, and single crystal elastic constants.
%RL,  as well as the elastic limit, and the elasto-plastic behavior, if plastic deformation takes place during the diffraction experiment. 
Thus, failing to fulfill mechanical equilibrium in the solution may be unavoidable. In other instances, ignoring the latter allows for a simplified and faster solution process, where large-scale optimization is already challenging and computationally intensive. However, when the material properties are known, taking into account mechanical equilibrium clearly provides a more constrained solution, and may significantly improve the stress/strain fields determination, and consequently, the spatial resolution, reliability, and application range of 3DXRD methods. The elastic deformation solvers are typically iterative, and thus mechanical constraints could be employed in one of many ways: to correct the final stress field solution,  in every iteration step by enforcing equilibrium in the latest solution, or as an additional component in an optimization target function.
%RL\cite{reischig2020three}.
%RL{\color{orange} /!!! or merge this into the paragraphs below/}

%RL{\color{orange} /!!! by Peter: this paragraph needs review! This needs to be a new paragraph. Double check references! reischig2019 & reischig2020 are using near-field DCT, and only use grain averages as a starting point, and in reischig2020 for regularization, but they seek a proper voxelized solution by modelling all features of the diffraction spots which are deformed projections of the grains./}
%RLThis is due to grain averaging (i.e., approximating the stress field by a piecewise constant field using the grain averages inside a tessellated microstructure generated from the centers of mass of the grains) in the original methodology and the lack of micromechanical constraints as part of the optimization procedure, in the recent, more advanced, ff data inversion methods \cite{shen2020voxel}.{\color{orange}  /!!! these references were wrong/.}

Given the aforementioned problem, an approach to solve the lack of equilibrium of the measured stress field is to combine micromechanical modelling and simulations with experimental observations. In different contexts, this combination has been recently done by Pokharel and Lebensohn \cite{pokharel2017instantiation} using a fast Fourier transform (FFT)-based elasto-viscoplastic (EVPFFT) model, and by Chatterjee {\it et al.} \cite{chatterjee2017study} adopting a field dislocation mechanics finite element model, to complement X-ray diffraction based experimental information and characterize residual and internal stresses. Also, Pagan and Beaudoin \cite{pagan2019utilizing} used lattice orientation and crystal plasticity kinematics to recover the geometry, and further calculated the stress through finite element simulation. 

While the above authors solve physics-based field equations, McNelis, Dawson and Miller \cite{mcnelis2013two} proposed a new approach where they seek to impose equilibrium to match the lattice stresses inferred from XRD and elasticity.  In particular, they were interested in large mechanical parts (at the order of  meters) where the detection spots are limited.   Therefore, they proposed a two-scale method, where the continuum scale stress field imposes equilibrium to match the lattice scale stresses.   The approach was extended to three dimensions by Demir {\it et al.} \cite{demir2013computational}, and has found successful applications \cite{park2013quantifying, park2018non}. 
However, this approach does not impose equilibrium at the grain and sub-grain scale, which is the focus of our work.

%The problem of stress reconstruction becomes especially hard for large mechanical parts (at the order of meters) since the detection spots are limited.  
%
%proposed a two-scale method, where the continuum scale stress field imposes equilibrium to match the lattice scale stresses .   The approach was extended to three dimensions by Demir {\it et al.} \cite{demir2013computational}, and has found successful applications \cite{park2013quantifying, park2018non}. 
%We note that this method does not solve the full-field constitutive equations but only imposes stress equilibrium and boundary conditions on the stress field {\color{blue} The last sentence is unclear}.

This paper presents a novel methodology to impose micromechanical constraints, i.e. stress equilibrium at the subgrain and grain scale, to an non-equilibrated voxelized stress field obtained, for example, by means of synchrotron X-ray diffraction techniques.  The main idea is to find the equilibrated stress field closest (in $L^2$-norm sense) to a measured (and possibly) non-equilibrated stress field, via the solution of an optimization problem based on the Hodge decomposition of a symmetric 2nd rank tensorial field.  The Hodge decomposition is a generalization of the classical Helmholtz decomposition that states that any sufficiently smooth vector field can be decomposed into the sum of an irrotational vector field and a solenoidal vector field.  We use the version of the Hodge decomposition for symmetric tensorial fields developed by Geymonat and Krasucki \cite{geymonat2009hodge}.  We present an efficient numerical implementation of the Hodge decomposition of symmetric 2nd rank tensorial fields using FFTs, and apply it to the analysis/determination of equilibrated stress fields in polycrystalline materials, including 3-D space-resolved stress measured by X-ray diffraction. As mentioned before, these stress fields are in general not divergence-free, due to piecewise approximation based on grain averages and/or absent the consideration of any micromechanical constraint in the data inversion procedure. Using the proposed decomposition and formulating an optimization problem, unbalanced stress fields can be filtered to extract their divergence-free part. The proposed methodology has been further modified/extended to deal with interfaces and discontinuities occurring at grain boundaries.  We begin by demonstrating that the method is capable of good recovery of synthetic data obtained in the elastic regime with an FFT-based micromechanical model that provides an objective ground-truth (a validation methodology originally used in \cite{shen2020voxel}), and then apply it to  a stress field actually obtained from diffraction data.  

The outline of the paper is as follows. In Section 2 we present a summary of the 3D orientation and strain mapping approach developed by Reischig and Ludwig \cite{reischig2019three}, and its application to Gum Metal, a beta-Ti-based alloy, deformed and measured in-situ within the elastic regime. In Section 3 we recall the fundamentals of the Hodge decomposition and formulate the optimization problem that allow us to extract the divergence-free part of a general non-equilibrated stress field. In Section 4, we present the FFT-based method that enables an efficient numerical resolution of the previously formulated optimization problem when applied to a voxelized field, the type of data structure that naturally results from data reduction of 3DXRD experiments. Section 5 shows applications of the proposed method to synthetic piecewise constant stress fields with a known ground truth, and stress fields in Gum Metal measured by 3DXRD. In both cases, we show that the largest corrections are obtained near grain boundaries.  In Section 6, we draw conclusions and give perspectives of the adoption of the proposed method in 3DXRD data inversion packages.

%In this work, we propose a simple and general method based on experimental stress data. To filter and smooth the noisy data obtained from experiments, Our method imposes the equilibrium constraint in a similar spirit for 3D stress field. We project the stress field, a $3\times3$ symmetric tensor field, into a equilibrated stress field, a divergence free function space. Hodge decomposition of symmetric matrix field is proposed\cite{geymonat2009hodge}. We derived the corresponding projection operator for periodic medium, which leads to a biharmonic equation. The equation is solved using an FFT based method and further modified to deal with stress discontinuity. The projection method is tested with several synthetic data, proving a good recovery of the original data. We also applied it to a high energy X-ray diffraction experiment, showing its ability to deal with no-periodic situations. Although stress measurement from X-ray diffraction serves our motivation, the method is general and able to be implemented in post-processing of other experimental data.

\section{Experimental characterization of elastic strain tensors} \label{sec:expt}

% /!!! Peter: It seems arbitrary to discuss the Shen2020 method here after a long introduction, but not the other methods mentioned above, so I deleted and recycled this./  
%Recently two different approaches for reconstruction of local intragranular strain and orientation from near-field X-ray diffraction data have been proposed and applied to experimental data \cite{Shen2020,Reischig2020}. Both techniques use small rotation increments $(\le 0.05^{\circ})$ during acquisition and the components of the deformation gradient tensor are determined by minimizing the difference between experimentally observed and forward simulated diffraction signals. The first approach \cite{Shen2020} is an extension of the forward modeling technique \cite{Suter2009,Li2013} and has been demonstrated on data acquired in 1D (slice by slice) acquisition mode. The second one, named Iterative Tensor Field Reconstruction, is related to X-ray diffraction contrast tomography and estimates the components of the deformation gradient tensor via iterative solution of a linearized kinematical diffraction model. Here we briefly recall the principal steps and assumptions of this second technique which has been described in detail in \cite{Reischig2020}.\\

\subsection{Data acquisition}
The experimental dataset used in this work to demonstrate the proposed stress filtering method was recorded in a regular Diffraction Contrast Tomography (DCT) scan \cite{ludwig2009three,Reischig2013}. The method uses a parallel monochromatic synchrotron X-ray beam and records diffraction spots from individual grains of the polycrystalline sample on a near-field area detector while the sample is rotated continuously over $360^{\circ}$. The image stack is subject to a number of pre-processing steps, at the end of which the diffraction spots are segmented and indexed according to their grain of origin and their $(hkl)$ Miller indices. The indexing is based on crystallographic principles and the knowledge of the crystal structure and approximate undeformed lattice cell parameters of the one (or possible several) phases present in the polycrystal. Each diffraction spot is stored as a $(u,v,w)$ volume array representing the diffracted intensity distribution as a 3D scalar field: $u$ and $v$ being the horizontal and vertical image coordinates, and $w$ is the rotation angle. The use of monochromatic X-rays and a high-resolution  detector provides sensitivity to the local orientation and local unit cell parameters within the crystallites. Local variations in lattice parameters, i.e. a change of the shape and dimensions of the unit cell, in this case is interpreted as mechanical elastic strain, and any effects from a possible local change of chemical composition is neglected. 

The results presented in this study were obtained from a polycrystalline Gum Metal sample with a composition of Ti-36Nb-2Ta-3Zr-0.3O $wt\%$ and mean grain diameter of $61\mu$m, from the "low load" scan presented in \cite{reischig2020three}. Gum Metal is able to sustain elastic strains up to $2\%$ elongation and above, thus its nickname. A $500\mu$m tall gauge volume of a tensile specimen with an approximately $600\mu$m wide rectangular cross-section was illuminated in its entirety with a box beam during the $360^{\circ}$ rotation, using 40 keV beam energy, 1.5 sec exposure time per image, $0.05^{\circ}$ angular step size, resulting in 7200 images in total. The detector pixel size was $1.4\mu$m, and the rotation axis to detector distance was 7 mm. The sample was mounted in a miniature tensile rig \cite{Gueninchault2016} and a DCT scan at a uniaxial external tensile stress of 34 MPa was acquired. On average 29 diffraction spots from the first three families of reflections could be recorded for each of the $\sim1430$ grains in the illuminated sample volume, which served as the input data for the reconstruction.

\subsection{Reconstruction of the sub-grain elastic deformation field}
The first stage of the data analysis follows the now standard DCT processing route, with a calibration procedure based on the dataset itself \cite{reischig2019three, reischig2020three}. The centroid positions and mean orientations of the grains were determined based on their diffraction spot centroid metadata. The initial 3D shapes of the grains were reconstructed iteratively from the diffraction spots using the SIRT algorithm, and assuming a constant orientation and elastic strain distribution throughout the grain volume.

The deformation analysis applied here corresponds to an earlier version of the novel Iterative Tensor Field (ITF) Reconstruction method described in detail in \cite{reischig2020three}, the main difference being that here no Tikhonov regularization term was applied in the elastic deformation optimization target function (Equation (39) in  \cite{reischig2020three}). The ITF method was used to retrieve the complete local elastic strain tensor (6 parameters), the local lattice misorientations from a reference grain orientation (3 parameters) and refine the grain boundaries over the entire 3D voxelated gauge volume in the specimen. ITF is based on a vector representation of the 3D distribution of X-ray diffracting powers of the crystallites $\mathbf{p}$ (one scalar per voxel) and a 3D intragranular deformation field $\mathbf{d}$ (9 deformation components per voxel) of their crystal lattice, and aims to reconstruct this $\mathcal{F}(x,y,z): \mathbb{R}^3 \mapsto \mathbb{R}^{1+9}$ tensor field on a grain-by-grain basis in a static deformed state, measured in a single scan. Mosaicity, i.e. a variation of crystal lattice orientations without associated stress, is allowed in the solution, hence the compatibility equations are not enforced. A diffraction spot is effectively treated as a projection of the 3D grain shape into the 3D detector domain $(u,v,w)$, where the sub-grain deformations (local misorientations and elastic strains) determine the projection geometry that is unknown. ITF builds on the principles of kinematical diffraction and ray tracing, and the solution for the elastic deformation field is found by minimizing the differences between the measured pixel intensities of the detector $\mathbf{q}^{mes}$ (ordered as a column vector) and the simulated intensities $\mathbf{q}$ in the diffraction spots of a grain in an iterative large scale optimization. This is a strongly non-linear and usually under-determined problem but in each iteration step a local linear problem at the latest solution can be formulated in matrix form:
\begin{flalign}\label{eq:linprob1}
\Delta\mathbf{q} = \mathbf{q}^{mes} - \mathbf{q} =
\boldsymbol{A}_{d} \, \Delta\mathbf{d}
\end{flalign}
where $\Delta\mathbf{d}$ (a column vector) contains the corrections sought to be applied to each elastic deformation component of each active grain voxel in the current iteration step, either positive or negative. The $\boldsymbol{A}_{d}$ matrix contains the intensity contributions to each pixel (one row per pixel) from each elastic deformation component correction of the active grain voxels. The $\boldsymbol{A}_{d}$ matrix is non-sparse, and it needs to be recomputed frequently due to the strong non-linearity of the model. Translated into a least-squares problem, the solver minimizes the objective function $\Gamma$ which in this case is the square of the L2 norm (denoted by $\left\|.\right\| _2$) of the residual vector:
\begin{flalign}\label{eq:}
\Gamma =
\left \| \Delta\mathbf{q} - \boldsymbol{A}_d \Delta\mathbf{d} \right \| _2^2 
\end{flalign}
\begin{flalign}
\Delta\mathbf{d}^* = \underset{\Delta\mathbf{d}}\argmin(\Gamma)
\end{flalign}
The elastic deformation corrections $\Delta\mathbf{d}^*$ found in the latest iteration step are applied to the current elastic deformation field of the grain. A 3D smoothing operation is then performed on the elastic deformation field in each step, which effectively acts as regularization in the optimization and results in a more realistic field. This smoothness suppresses but does not eliminate the deviations from local mechanical equilibrium across a grain volume.

The corrections to the diffracting powers, which determine the grain shapes, are recomputed in a separate step in a similar way but less frequently, using the latest elastic deformation field (i.e. the latest projection geometry). 

Errors in the elastic deformation fields result from a number of sources in the acquisition and data processing. The strain sensitivity here is expected to be in the high single digits of $10^{-4}$. The overall deformation sensitivity of the setup was limited by the angular step size, pixel size and detector distance. The relatively small number of diffraction spots meant the problem was more ill-posed. Some systematic deviations in intensity were caused by the simplifications in the model of the image formation process. Random noise from the photon counting statistics, fluorescence and scattering processes was often significant. The grain-by-grain reconstruction approach is inherently prone to higher errors near grain boundaries. These aspects and potential future improvements of the acquisition and data processing, including the handling of all grains simultaneously and ways to include the local mechanical equilibrium, are described in more detail in \cite{reischig2020three}. The fit quality in the ITF deformation solver can be monitored through the residual. In this study, the fitting is considered to have failed in several percent of the grains, mainly at the top and bottom of the gauge volume and at the free surface. Although the dataset does not constitute the state-of-art measurement capabilities, it is perfectly suitable to demonstrate the proposed stress filtering method.

\section{Proposed Method}
\subsection{Hodge decomposition}

We use the following notation.  Let $\Omega \subset {\mathbb R}^3$ denote a rectangular domain and $H^1_{per}(\Omega, {\mathbb R})$ be the set of all functions $u \in H^1_{loc} ({\mathbb R}^3,{\mathbb R})$ that are periodic with period $\Omega$.  The Helmholtz decomposition is a fundamental theorem in vector calculus that has applications in many fields including elasticity, incompressible flows and electromagnetism (see for example,  \cite{cantarella2002vector}).  
\begin{theorem}
Given any vector field $v \in L^2(\Omega, {\mathbb R}^3)$, there exists a scalar field $\varphi \in H^1_{per}(\Omega, {\mathbb R})$ and a vector field $w \in H^1_{per}(\Omega, {\mathbb R}^3)$, such that,
\begin{equation*}
    v = \nabla \times w + \nabla \varphi, \quad \nabla \cdot w = 0.
\end{equation*}
Further $\varphi, w$ are unique up to a constant.
\end{theorem}
The Hodge decomposition is a generalization to tensor fields, and we use a version due to Geymonat and Krasucki \cite{geymonat2009hodge} for symmetric fields.  Let ${\mathbb M}^3$ denotes the linear space of all second-order matrices and ${\mathbb M}_{sym}^3$ the linear space of all second-order symmetric matrices.
\begin{theorem}
Given a symmetric matrix field $A \in L^2(\Omega, {\mathbb M}^3_{sym})$, there exist a vector field $y \in H^1_{per}(\Omega, {\mathbb R}^3)$, a symmetric matrix field $H \in H^2(\Omega, {\mathbb M}^3_{sym})$ and a constant matrix field $c_0$ such that,
\begin{equation*}
    A = \nabla \times H \times \nabla + \frac{1}{2}(\nabla y + \nabla y^{T} ) + c_0, \quad \nabla \cdot H = 0
\end{equation*}
where $(\nabla \times H \times \nabla)_{kn} = \epsilon_{ijk} \epsilon_{lmn} \nabla_i \nabla_l H_{jm}$ in indicial notation.  
Further $y, H$ up to a constant, and $c_0$ are unique.
\end{theorem}

\subsection{Problem formulation}
Given a symmetric matrix field $S_{exp} \in L^2(\Omega, {\mathbb M}^3_{sym})$, we want to find the closest field $S_{equil} \in H^1(\Omega, {\mathbb M}^3_{sym})$ in $L^2$ norm, such that $\nabla \cdot S_{equil} = 0$. In other words,
\begin{equation*}
    S_{equil} = \argmin \left\{ ||S - S_{exp}||_2:  S \in H^1_{per} (\Omega, {\mathbb M}^3_{sym}),  \nabla \cdot S = 0 \right\}.
\end{equation*}
It is easy to verify using the divergence theorem that $\langle(\nabla y + \nabla y^T), S\rangle_{H^1} = 0$ for any $y \in H^1_{per}(\Omega, {\mathbb R}^3)$ and $S \in H^1_{per}(\Omega, {\mathbb M})$ that satisfies $\nabla \cdot S = 0$.  Since $\nabla \cdot S_{equil} = 0$, it follows that the Hodge decomposition of $S_{equil}$ has $y = 0$, or
\begin{equation} \label{eq:sequil}
    S_{equil} = \nabla \times H \times \nabla + c_0, \quad \nabla \cdot H = 0
\end{equation}
for some $H \in H^2_{per}(\Omega, {\mathbb M}^3_{sym}) $ and $c_0 \in {\mathbb M}^3_{sym}$.  It follows that
\begin{equation}
    S_{exp} = S_{equil} + \frac{1}{2}(\nabla y + \nabla y^{T} ) = \nabla \times H \times \nabla + c_0 + \frac{1}{2}(\nabla y + \nabla y^{T} ).
\end{equation}
Now, taking the average over $\Omega$ and again using the divergence theorem, $c_0 = \overline{S}_{exp}$ where we use the overhead bar to denote spatial average $\overline{A} = \frac{1}{|\Omega|} \int_{\Omega} A dX$.  Putting all this together, the original minimization problem can then be written as
\begin{equation}     \label{eqnOptH}
\begin{split}
H = \argmin \left\{{\mathcal L}(A) :  A \in H^2_{per}(\Omega, {\mathbb M}^3_{sym}), \nabla \cdot A = 0 \right\}  \\
\quad \quad \quad \text{where}  \quad {\mathcal L}(A) := ||\nabla \times A \times \nabla - \widetilde S_{exp}||_2.
\end{split}
\end{equation}
where tilde denotes the demeaned part $\widetilde{S}_{exp}=S_{exp}-\overline{S}_{exp}$. We use this variational problem to find $S_{equil}$.

The first variation
\begin{equation*}
    \delta {\mathcal L} = \int_{\Omega} (\nabla \times H \times \nabla - \widetilde S_{exp}) \cdot (\nabla \times \delta H \times \nabla) \ dX = 0
    \quad \forall \ \delta H \in {\mathcal A}
\end{equation*}
where ${\mathcal A} = \{A \in H^2_{per}(\Omega, {\mathbb M}^3_{sym}): \nabla \cdot A = 0 \}$.
Notice that for any $A,B \in H^1_{per}(\Omega, {\mathbb M}^3_{sym})$,
\begin{equation*}
    \int_{\Omega} (\nabla \times A) \cdot B - A \cdot (\nabla \times B) \ dX = \int_{\Omega} \nabla \cdot (A \times B) \ dX = 0.
\end{equation*}
where the second equality follows from the divergence theorem.  Thus,
\begin{equation*}
    \int_{\Omega} (\nabla \times A) \cdot B \ dX = \int_{\Omega} A \cdot (\nabla \times B) \ dX.
\end{equation*}
Applying this equality twice to the functional, we obtain
\begin{equation*}
    \delta {\mathcal L} = \int_{\Omega} (\nabla \times \nabla \times H \times \nabla \times \nabla - \nabla \times \widetilde S_{exp} \times \nabla) \cdot \delta H \ dX = 0
    \quad \forall \ \delta H \in {\mathcal A}.
\end{equation*}
We obtain the Euler-Lagrange equation, 
\begin{equation*}
    \nabla \times \nabla \times H \times \nabla \times \nabla = \nabla \times \widetilde S_{exp} \times \nabla.
\end{equation*}
Since $\nabla \times (\nabla \times H) = \nabla (\nabla \cdot H) - \nabla^2 H$ and $\nabla \cdot H = 0$,  the Euler-Lagrange equation simplifies to 
\begin{equation}
    \nabla^4 H = \nabla \times \widetilde S_{exp} \times \nabla.
    \label{eqnELH}
\end{equation}
We solve this equation to find $H$ and then $S_{equil}$.  Recall that $H$ is unique up to the constant, but the constant does not affect $S_{equil}$: so we choose the constant to make $\overline{H} = 0$.

%In the end, we want to emphasize that this minimizer $H$ is unique up to a constant, given the periodic boundary condition. Since the biharmonic equation could be reformulated as two Poisson equation, both of which has a periodic condition.
%\begin{equation*}
%    \begin{cases}
%      & \Delta H = G\\
%      & \Delta G =  \nabla \times \widetilde S_{exp} \times \nabla\\
%    \end{cases}       
%\end{equation*}
%We note that the solution of periodic Poisson equation is unique up to a constant. So $G$ is unique up to a constant, as it is a solution of the second equation. But we also notice that $\Delta H = G$ enforce $\overline{G} = 0$ by integration over two sides. Thus $G$ is unique. So $H$ is unique up to a constant as $H$ is a solution to the first equation. Further, this constant does not effect the value of $S_{equil}$. We could set $\overline{H} = 0$ to resolve non-uniqueness.

\subsection{Solution strategy}

Given $ S_{exp}$, we average over the volume to find $c_0$ and $\widetilde S_{exp}$.  We then solve Eq. ($\ref{eqnELH}$) to find $H$.  Finally, we use Eq. (\ref{eq:sequil}) to obtain $S_{equil}$.  The solution of Eq. (\ref{eqnELH}) poses a challenge since it involves the fourth order biharmonic equation for $H \in H^2$, which requires continuous functions and derivatives.  This in turn requires either higher order elements using finite-element discretization, or higher order differences in finite-difference methods. In either situation, it gives rise to stiff numerical problems \cite{argyris1976finite,gupta1979direct}. Another approach is to break it up into two harmonic problems and to solve them iteratively \cite{cheng2000some}. Here, we solve Eq. (\ref{eqnELH}) using Fourier transforms. This poses two challenges.  The first is that the experimental stress-field $\ref{eq:sequil}$ is not periodic. To overcome this, a buffer is filled around the voxelized representation of the material to make the domain a periodic box. The properties of the buffer near the lateral surfaces are chosen to give zero stress. We refer to a detailed discussion in section \ref{sec:expt}. Second, the stress may suffer discontinuities across grain boundaries, interfaces and across the boundaries of the periodic unit cell. A Fourier series does not converge to discontinuous functions in the $L^\infty$ norm, and this in turn leads to ringing artifacts in practice.  We address this in the next section.

\section{Different operators in Fourier space}

Let $\Omega = (0,L)^3$ to be a cube and let us discretize it with a $N \times N \times N$ uniform cubic grid for $N$ even.  The corresponding domain in the Fourier space is $(- \pi/L, \pi/L)^3$ and again discretized uniformly with a $N \times N \times N$ grid.   For any $f: \Omega \to {\mathbb R}^3$, the discrete Fourier transform (DFT) $\hat{f}$ satisfies
\begin{equation}
    f(i,j,k) = \sum_{l = 1}^N \sum_{m = 1}^N \sum_{n = 1}^N \hat{f}(l,m,n) exp(\frac{2 \pi \mathrm{i}}{L} ((i-1)(l-1) + (j-1)(m-1) + (l-1)(k-1)),
    \label{eqnFFT}
\end{equation}
\begin{equation}
    \hat{f}(l,m,n) = \sum_{i = 1}^N \sum_{j = 1}^N \sum_{k = 1}^N f(i,j,k) exp(\frac{-2 \pi \mathrm{i}}{L} ((i-1)(l-1) + (j-1)(m-1) + (l-1)(k-1)),
    \label{eqniFFT}
\end{equation}
where $\mathrm{i} = \sqrt{-1}$ (distinct from index $i$).

\subsection{Continuous differential operator}
The continuous differential operator (CDO) of $f$ in Fourier domain is obtained point-wise multiplication. For example, the partial derivative of f with respect to x is
\begin{equation} \label{eq:cdo}
    \frac{\partial f(i,j,k)}{\partial x} = \sum_{l = 1}^N \sum_{m = 1}^N \sum_{n = 1}^N \mathrm{i} \xi_x \hat{f}(l,m,n) exp(\frac{2 \pi \mathrm{i}}{L} ((i-1)(l-1) + (j-1)(m-1) + (l-1)(k-1))
\end{equation}
where
\begin{equation}
    \xi_x = 
    \begin{cases}
    &\frac{2 \pi}{L} l, \quad \text{if } 1 \leq l < \frac{N}{2}\\
    &0, \quad \text{if } l = \frac{N}{2}\\
    &\frac{2 \pi}{L} (l - N), \quad \text{if } \frac{N}{2} < l \leq N
    \end{cases}
\end{equation}
Higher order derivatives are a direct composition of first derivatives, with modification on the coefficients of the highest frequency term (see for example \cite{johnson2011notes}).

\subsection{Discrete differential operator}
We are interested in studying situations with grain boundaries and interfaces across which the stress may suffer discontinuities.  A Fourier series does not converge to such a function in the $L^{\inf}$ norm even as $N \to \infty$.  In practice, we encounter ringing artifacts or spurious oscillations at finite $N$.   A solution to this problem is to construct the differential operator by taking finite differences and then using DFT  \cite{berbenni2014numerical,vidyasagar2017predicting}. The first order central difference approximation reads,
\begin{equation} \label{eq:diff}
    \frac{\partial f(i,j,k)}{\partial x} = \frac{f(i+1,j,k) - f(i-1,j,k)}{2 \Delta h}
\end{equation}
%\begin{equation*}
%    \frac{\partial f(i,j,k)}{\partial y} = \frac{f(i,j+1,k) - f(i,j-1,k)}{2 \Delta h} 
%\end{equation*}
%\begin{equation*}
%    \frac{\partial f(i,j,k)}{\partial z} = \frac{f(i,j,k+1) - f(i,j,k-1)}{2 \Delta h}
%\end{equation*}
etc., and therefore the discrete differential operator is given by Eq. (\ref{eq:cdo}) with the substitution
\begin{equation*}
    i\xi_x \leftarrow \frac{i}{\Delta h} \sin(\frac{2\pi(m - 1)}{N}), %\quad
%    i\xi_y \leftarrow \frac{i}{\Delta h} \sin(\frac{2\pi(n - 1)}{N})\quad 
%    i\xi_z \leftarrow \frac{i}{\Delta h} \sin(\frac{2\pi(l - 1)}{N})
    \label{eqnDisOperator}
\end{equation*}
and analogously for the other partial derivatives. For example, 
\begin{equation*}
    (\nabla \times H \times \nabla)_{kn} = \epsilon_{ijk} \epsilon_{lmn} \nabla_i \nabla_l H_{jm},
\end{equation*}
we substitute $\nabla_x$ with $\frac{i}{\Delta h} \sin(\frac{2\pi(m - 1)}{N})$ for the discrete approximation.
Note that this first order operator approximates the continuous differential operator for small $\xi_x$, but then decays thereby acting as a low-pass filter. This resolves the overshooting and ringing at the interfaces.  However, the convergence with respect to mesh size is reduced to second order from the exponential convergence of the CDO.

The biharmonic operator adopts the following 25 point stencil that is consistent,
\begin{align*}
    \nabla^4 f(i,j,k) &= 42f(i,j,k) - 12(f(i-2,j,k) + f(i+2,j,k) + f(i,j-2,k) + f(i,j+2,k)\\
                      &+ f(i,j,k-2) + f(i,j,k+2)) + f(i-4,j,k) + f(i+4,j,k) + f(i,j-4,k)\\
                      &+ f(i,j+4,k) + f(i,j,k-4) + f(i,j,k+4) + 2(f(i-2,j-2,k)\\
                      &+ f(i-2,j+2,k)  + f(i+2,j-2,k) + f(i+2,j+2,k) + f(i,j-2,k-2)\\
                      &+ f(i,j-2,k+2) + f(i,j+2,k-2) + f(i,j+2,k+2) + f(i-2,j,k-2)\\
                      &+ f(i-2,j,k+2) + f(i+2,j,k-2) + f(i+2,j,k+2)).
\end{align*}
Notice that this is based on the second neighbor to make it consistent with the first order difference , Eq. (\ref{eq:diff}).
We obtain the corresponding operator in Fourier space from the usual Fourier space biharmonic operator with the substitution
\begin{align*}
    \xi_x^4+\xi_y^4+\xi_z^4 &\leftarrow (42 - 24 (cos(4\pi (m-1)/N) + cos(4\pi (n-1)/N)) + cos(4\pi (l-1)/N))\\
                            &+ 2 (cos(8 \pi (m-1)/N) + cos(8 \pi (n-1)/N) + cos(8 \pi (l-1)/N))\\
                            &+ 8(cos(4 \pi (m-1)/N)cos(4 \pi (n-1)/N) + cos(4 \pi (n-1)/N)cos(4 \pi (l-1)/N)\\ 
                            &+ cos(4 \pi (m-1)/N)cos(4 \pi (l-1)/N))) \frac{1}{16\Delta h^4}.
\end{align*}

\section{Numerical Examples}
We study two examples, one using synthetic data with a known ground truth and another with experimental data.  

\subsection{Synthetic data with known ground truth}

\begin{figure*}
   	\begin{center}
   		\resizebox{.6\textwidth}{!}{%
   			\includegraphics[height=3cm]{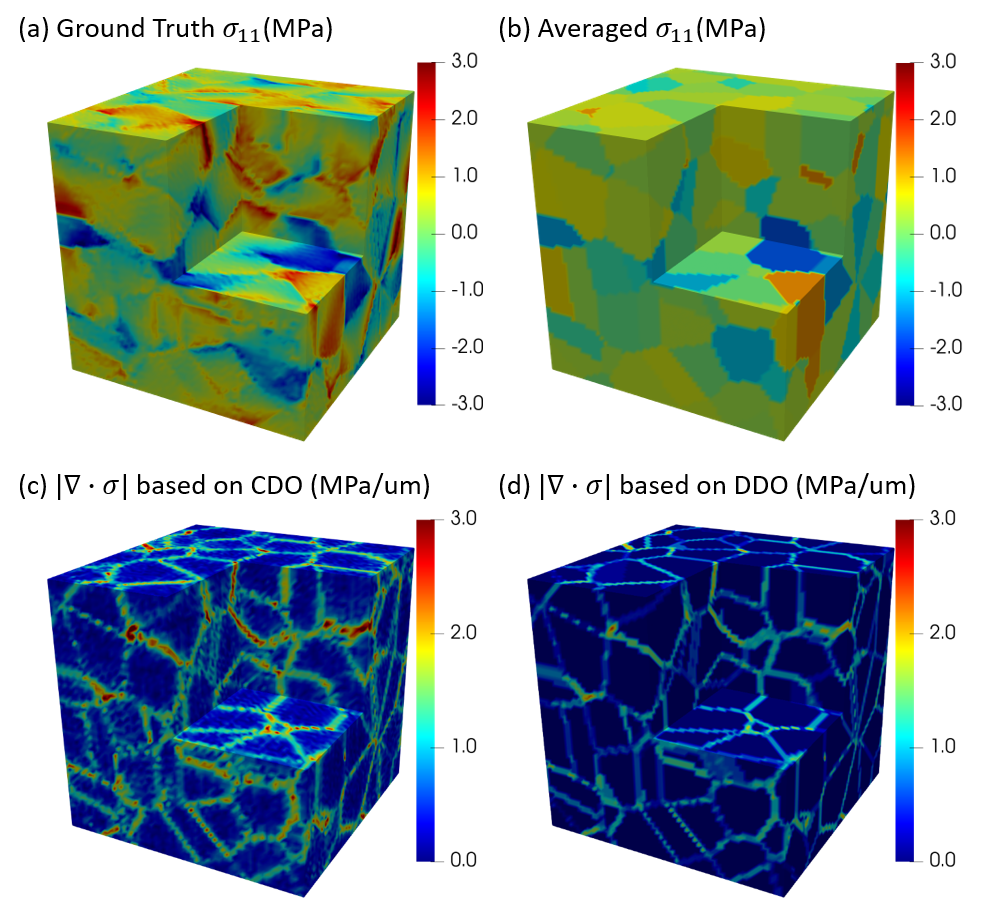}
   		}
   	\end{center}
   	\caption{Synthetic data: (a) lateral stress field $\sigma_{11}$ obtained with FFT-based model; (b) grain averaged $\sigma_{11}$; (c) divergence based on CDO; (d) divergence based on DDO. }
   	\label{fig_synthetic}
\end{figure*}

We begin with synthetic data from the results of a simulation using the FFT-based method \cite{lebensohn2012elasto} to obtain an equilibrated stress field on a periodic unit cell representing a Cu polycrystal deformed in tension in the elastic regime. This polycrystalline unit cell is discretized by a $64^3$ grid on a $[0,64]  \mu m^3$ box, and contains 100 Cu single crystal grains generated by periodic Voronoi tesselation. The Cu single crystal elastic constants reflect their cubic anisotropy, with $C_{11}= 168.4$ GPa, $C_{12}=121.4$ GPa and $C_{44}=75.4$ GPa, which determines heterogeneous, piece-wise constant elastic properties associated with the different crystal orientations, and, therefore, a non-uniform stress field, when the polycrystal is subjected to load in the elastic regime. For this analysis, the unit cell was loaded to a strain of $10^{-5}$ in uniaxial tension along axis $x_3$, corresponding to a longitudinal stress $\Sigma_{33}=6.32$ MPa, and zero lateral stresses: $\Sigma_{11}=\Sigma_{22}=0$, applied to the unit cell. Figure \ref{fig_synthetic}(a) shows the $\sigma_{11}$ component of the local stress field (in MPa), which fluctuates with respect to the macroscopic value $\Sigma_{11}=0$.  Since this synthetic data was generated  using a physically meaningful model that uses CDO to solve the governing equations of micro-mechanics (stress equilibrium and strain compatibility), the resulting stress field is indeed equilibrated (and thus divergence-free with respect to the CDO).

We start with this equilibrated stress field and obtain a {\it grain-averaged} stress field by averaging the stress over the grains -- Figure \ref{fig_synthetic}(b) shows the $\sigma_{11}$ component of the grain-averaged stress.  This piece-wise constant field is not divergence-free, and the divergence is concentrated at the grain boundaries.  We compute the divergence of the grain-averaged stress using both CDO and DDO -- the results are shown in Figure \ref{fig_synthetic}(c) and Figure \ref{fig_synthetic}(d) respectively.  We observe that the divergence is generally concentrated at the grain boundaries, but CDO smears it into the grain interior.  This is a result of the spurious oscillation and overshooting that is characteristic of CDO on discontinuous functions.

%We present our results with a comparison of two differential operators, CDO and modified DDO. The test data are from simulation results of EVPFFT model \cite{lebensohn2012elasto}. Fig.\ref{fig_synthetic} (a) shows $\sigma_{11}$ distribution of EVPFFT simulation results. This EVPFFT simulation is also a FFT-based method, and it applies first order continuous differential operator while dealing with derivatives. The sample data is a box of dimension $[0,1]^3$ with mesh $64 \times 64 \times 64$. The stress field is balanced (i.e. divergence free w.r.t. first order continuous differential operator). We first take the stress in each grain to be its grain average (Fig.\ref{fig_synthetic} (b)). Thus inside each grain, the distribution is constant and divergence should zero. But at the grain boundaries, there should be a stress concentration. Fig.\ref{fig_synthetic} (c)(d) shows the norm of divergence of grain averaged distribution with two different operators. They both shows stress concentration at grain boundaries. CDO has a higher peak value compared with DDO. Since CDO overshoots at the discontinuities. CDO also smears the peak into the bulk grains, where the divergence should be zero. On contrast, DDO has clean separation between grains and boundaries.

\begin{figure*}
   	\begin{center}
   		\includegraphics[width=6in]{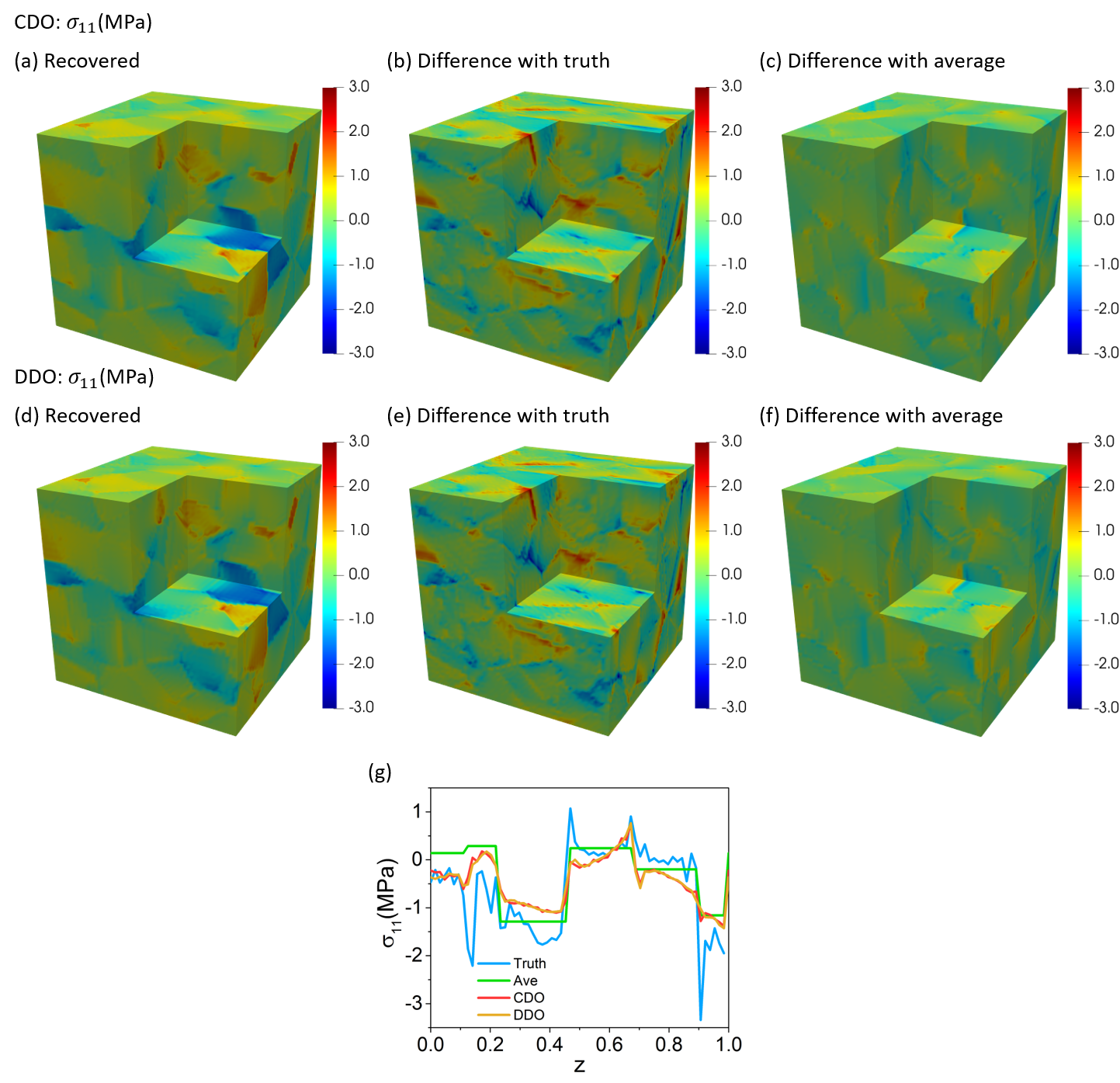}\\
   	\end{center}
   	\caption{Recovery from grain averaged data,  $\sigma_{11}$.  (a-c) Results of recovery with CDO: (a) recovered stress field, (b) difference between ground truth an recovered stress field and (c) difference between grain averaged and recovered stress field.   (d-f) Results of recovery with DDO: (a) recovered stress field, (b) difference between ground truth an recovered stress field and (c) difference between grain averaged and recovered stress field. (g) $\sigma_{11}$ along the line $0.5 \times 0.5 \times [0,1]$}
   	\label{fig_recover}
\end{figure*}

\begin{table}
    \begin{center}
         \begin{tabular}{c c c } 
         \hline
          & Diff with ground truth & Diff with average \\ [0.5ex] 
         \hline\hline
         CDO ($L^2$/$L^{\infty}$) & 0.0024/4.6041 & 0.0018/2.6945 \\ 
         \hline
         DDO ($L^2$/$L^{\infty}$) & 0.0025/4.6007 & 0.0018/2.6222 \\ [1ex] 
         \hline
        \end{tabular}
    \end{center}
\caption{Error in the recovery from grain averaged data}
\label{table_synthetic}
\end{table}

We now study if it is possible to obtain the original stress field from the grain-averaged data using our method.  We apply the proposed algorithm to the grain-averaged stress field using both CDO and DDO.  The results are shown in Figure \ref{fig_recover}.  We see that the recovery is imperfect, and this is confirmed in Table \ref{table_synthetic} which lists the $L^2$ and $L^\infty$ norms of the error ($|f(x)|_{L^{\infty}} = sup\{f(x) \text{$\forall$ $x$ in $\Omega$} \}$).  However, both CDO and DDO yield similar errors. In fact, Figure \ref{fig_recover}(g) shows the $\sigma_{11}$ component of the stress along the line $0.5 \times 0.5 \times [0,1])$.  Both operators recover similar fields though the CDO has more oscillations within the grains and  an overshoot at grain boundaries (see for example the grain boundary around $z=0.9$). 

To understand the error in recovery, recall that our algorithm only filters out the symmetrized gradient (curl-free) part and keeps all the divergence-free part of the given stress-field.  So, if the difference between the given data and the ground truth deviate by both curl-free and divergence-free fields, our algorithm will filter out the symmetrized gradient portion of the deviation but retain the divergence-free part of that deviation.  In our example, the difference between grain-averaged stress field and ground truth contains both curl-free and divergence-free components.  Our algorithm filtered out the former but not the latter and this is the error in both Figure \ref{fig_recover} and Table \ref{table_synthetic}.  We have verified this by taking the difference between grain-averaged stress field and ground truth, and applying our algorithm to it.   The resulting residual equals (close to machine precision) to the recovery error (difference between the ground truth and recovered) in both CDO and DDO.

\begin{figure*}[t]
   	\begin{center}
   		\resizebox{.6\textwidth}{!}{%
   			\includegraphics[height=3cm]{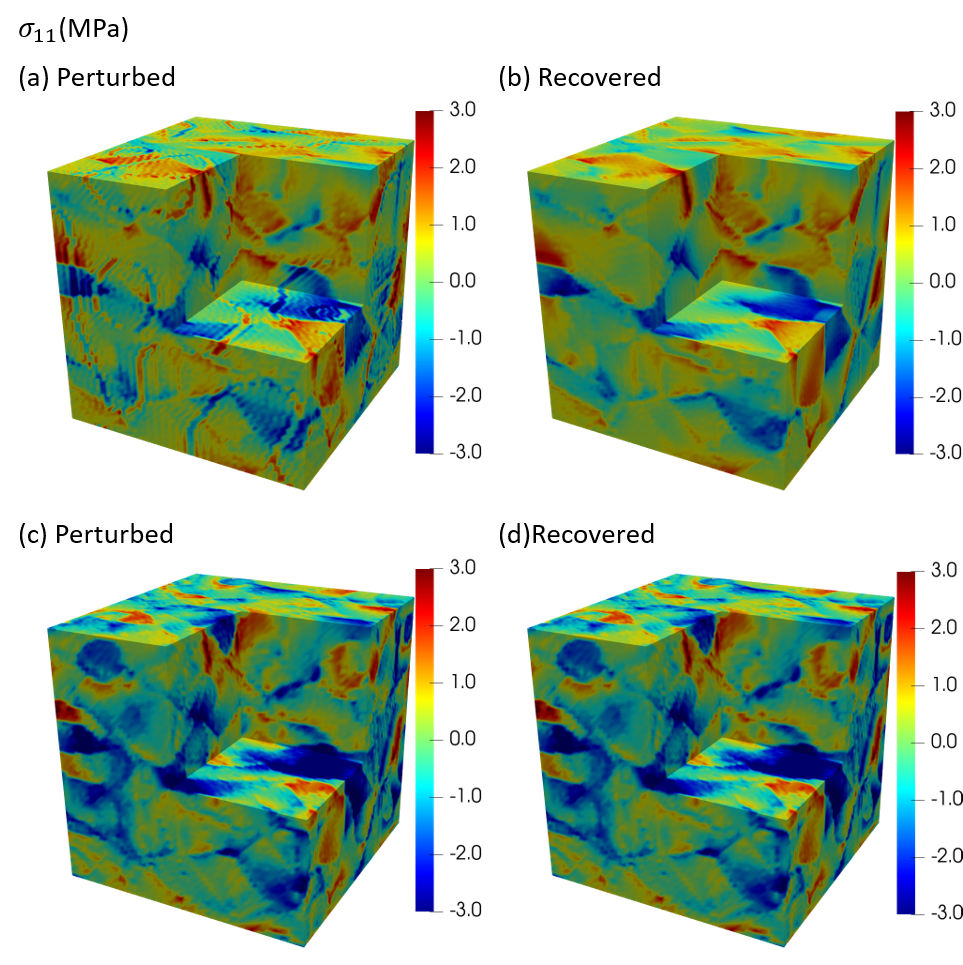}
   		}
   	\end{center}
   	\caption{(a,b) Recovery from a perturbation that is a symmetrized gradient (curl-free)  (a) perturbed and (b) recovered fields.  Compared to the ground truth in Figure \ref{fig_synthetic}(a), the recovery is very good -- $9.016 \times 10^{-6}$ in $L^2$ norm and $0.014$ in $L^{inf}$ norm.  (c,d) Recovery from a perturbation that is divergence-free (c) perturbed and (d) recovered fields.  The recovered field agrees with the perturbed field.}
   	\label{fig_perturb}
\end{figure*}

We further verify this in Figure \ref{fig_perturb}.  We perturb the ground truth by adding the symmetrized gradient $\nabla a + (\nabla a)^T$ of some vector field $a$ -- see Figure \ref{fig_perturb}(a). For a comparison of the scale, $|\sigma_{11}|_{L^{\infty}}$ of the original field is 6.02 MPa. $|\sigma_{11}|_{L^{\infty}}$ of the perturbation is 1.00 MPa. Our algorithm is able to recover the ground truth from the perturbation -- see Figure \ref{fig_perturb}(b) with minimal error: $9.016 \times 10^{-6}$ in $L^2$ norm and $0.014$ in $L^{inf}$ norm.  On the other hand, when we perturb the ground truth by adding a divergence free field (Figure \ref{fig_perturb}(c)), our algorithm returns the perturbed field and not the ground truth  (Figure \ref{fig_perturb}(d)). For a comparison of the scale, $|\sigma_{11}|_{L^{\infty}}$ of the original field is 6.02 MPa. $|\sigma_{11}|_{L^{\infty}}$ of the perturbation is 4.81 MPa. We note that whether the perturbation remains does not depend on the scale of perturbation.

In conclusion, our approach finds the best projection to divergence-free fields.  Therefore, it is able to filter out artifacts due to fields that are symmetrized gradients/curl-free, but unable to filter out artifacts that happen to be divergence-free.  Indeed, it is not possible to filter out divergence-free artifacts without additional knowledge in terms of material behavior.  For this reason, describing the average stress in each grain only provides limited information about the actual state of stress in the material.

\subsection{Experimental Sample}
\begin{figure*}
   	\begin{center}
   		\resizebox{.9\textwidth}{!}{%
   			\includegraphics[height=3cm]{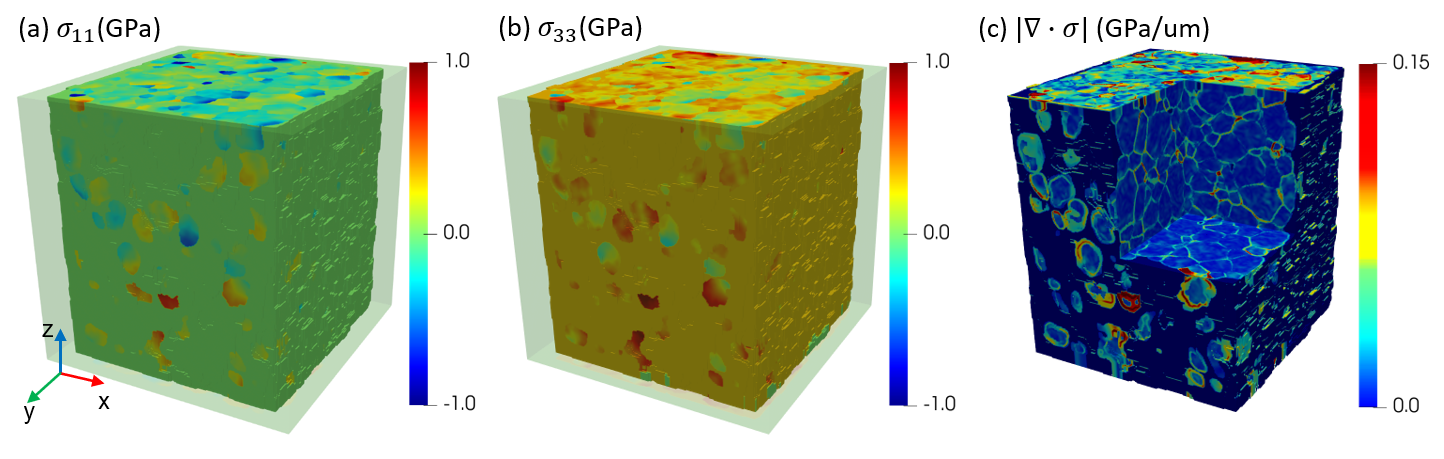}
   		}
   	\end{center}
   	\caption{Stress field as measured by DCT-ITF: (a) $\sigma_{11}$ (transparent box is the computational domain);  (b) $\sigma_{33}$.  (c) Magnitude of the divergence based on DDO. %{\color{orange} /Choose one of them, this one follows similar colormap//}
	}
   	\label{fig_exp}
\end{figure*}

In this section, we apply the proposed filtering methodology to the DCT experimental data on Gum Metal processed with the ITF method to obtain voxelized stress fields  as described in Section \ref{sec:expt}.   We focus on the measured stress field from a sample subjected to an average normal stress of 34 MPa along the $z$-axis while the other two lateral surfaces are stress free. In order to use our algorithm that assumes periodicity, we introduce a buffer region around the specimen. The buffer region on the side (along $x$-axis and $y$-axis) has zero stress to accommodate the free surface. We also note that we implicitly make the top and bottom to be periodically connected.
%There is a buffer region at the bottom, which is an extrusion of the stress state of the sample's bottom. Each pixel of the buffering region has the same stress state as the pixel above it. The introduction of this buffer region
% From Ricardo: I decided to eliminate the above, because the method can be numerically implemented with FFT packages (e.g. FFTW) that can handle arbitrary number of voxels, not just power-or-two, so I don't want to give the impression that the method is limited to fields with power-of-two numbers of voxels in each direction.
This spurious connection between top and bottom in the longitudinal direction evidently deteriorates the stress balance near those regions. However, we should note that the experimental measurement is also less reliable in those locations and in grains on the free surface of the sample that have high mosaicity as a result of the spark cutting process. This treatment still preserves the information in the bulk of the sample. The resulting unit cell of  $[0, 640\mu m]^3$, resolved at $256^3$ voxels is shown in Figure \ref{fig_exp}(a,b) along with the $\sigma_{11}$ and $\sigma_{33}$ components of the measured stress.  We note that the state of stress is quite heterogeneous.  Figure \ref{fig_exp}(c) shows the magnitude of the divergence of the measured stress-field calculated using DDO, and we observe high values of divergence at the grain boundaries and triple junctions.  This shows the the difficulty of accurate measurement of lattice strains and thus stresses where lattice structure is not preserved.   

\begin{figure*}
   	\begin{center}
   		\resizebox{.9\textwidth}{!}{%
   			\includegraphics[height=3cm]{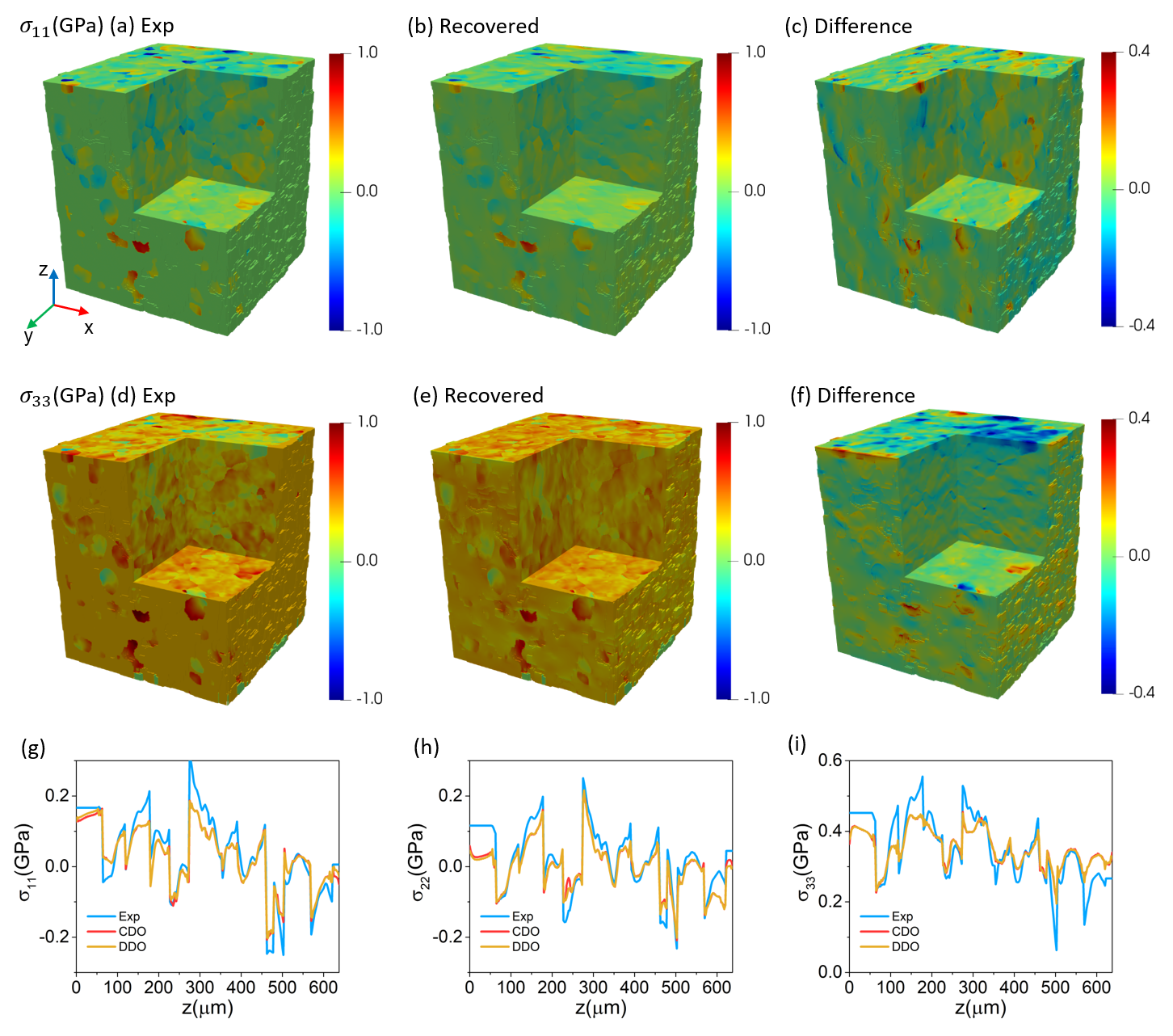}
   		}\\

   	\end{center}
   	\caption{Stress distribution: (a-c) $\sigma_{11}$; (d-f) $\sigma_{33}$; (g-i) $\sigma_{11}$, $\sigma_{22}$, $\sigma_{33}$ along the line $320 \mu m \times 320 \mu m \times [0, 640] \mu m$}
   	\label{fig_exp_recover}
\end{figure*}

We present the filtering results with DDO (Fig. \ref{fig_exp_recover}).   After filtering, the peak value is reduced. Most of the profile patterns are preserved. Comparing the difference, we notice that most of the region undergoes minor modification. Significant differences mostly occur around the grain boundaries. We also observe a significant correction of stress in the upper left grain, likely because the deformation solver failed for this grain and some other grains close to the surface. The main trends in Figure \ref{fig_exp_recover}(g) are well preserved though and most of the smaller features within the grains are retained, even if the stress magnitudes changed. This is an encouraging result, and suggests that the elastic deformation solver based on diffraction constraints returns realistic trends in the stress fields. The difference in the actual magnitudes of this measurement are somewhat less concerning, as these were less reliable due to the deformation resolution of the experimental setup not being fully adequate. The deformation sensitivity of the setup will be improved  in the future.

There are a number of outlier grains where the strain/stress state is extreme and significantly deviates from their neighbours. The ITF deformation fitting has potentially failed in these grains. A notable result of the filtering is that most of those grains are retained as outliers. Such high local deviations do not seem realistic, although their validity cannot be ruled out. Thus, it is not clear whether the filtered field is indeed more accurate in those regions. The measurement errors might be so high in those grains that the filtering cannot be expected to correct for them. 
%RL{\color{orange} / Please review and comment on this. /}       

A detailed comparison of the two operators is conducted by looking closely at the stress distribution along line  $320 \mu m \times 320 \mu m \times [0, 640] \mu m)$ (Fig. \ref{fig_exp_recover})(g).  Both methods yield similar results.  In some grains, the stress is not corrected much. But in the grain near $z \approx 300 \mu m$, there is a significant correction. The main difference between these two operators lie in grain boundaries. CDO and DDO both have some oscillations around the boundaries (e.g. $z \approx 250 \mu m$). But CDO provides a higher estimation at the boundaries. In contrast, DDO has a smoother transition but it is underestimating the values at the interface.

%A detailed comparison of different operators is conducted by picking up the stress distribution along line  $320 \mu m \times 320 \mu m \times [0, 640] \mu m)$ (Fig. \ref{fig_exp_com}). All three methods provide quantitatively similar results. In some grains, the stress is not corrected much. But in the grain near $z \approx 300 \mu m$, there is a significant correction. The difference between these two operators lie mainly near grain boundaries. CDO and DDO both have some oscillations around the boundaries (e.g. $z \approx 250 \mu m$). But they have a higher peak value. Especially CDO provides a higher estimation at the boundaries. On contrast, DDO has a smoother transition but it is underestimating the values at interfaces.
%\begin{figure*}
%   	\begin{center}
%   		\resizebox{.45\textwidth}{!}{
%   			\includegraphics[height=3cm]{./Figure/fig_exp_com.png}
%   		}
%   	\end{center}
%   	\caption{Comparison of stress distributions}
%   	\label{fig_exp_com}
%\end{figure*}

\section{Conclusions}
We developed a FFT-based Hodge decomposition method for symmetric matrix fields. To solve the problem of oscillations at interfaces and preserve the orthogonality of operators, we further developed a modified discrete differential operator.  We applied the Hodge decomposition to recover a equilibrated stress field from experimental stress measurements.  Since the method is based on a projection, it is able to filter out artifacts due to fields that are symmetrized gradients (curl-free), but unable to filter out artifacts that happen to be divergence-free.  Indeed, it is not possible to filter out divergence-free artifacts without additional knowledge in terms of material behavior.  We show that the algorithm can be effectively applied to experimentally obtained stress fields.  We have also investigated the difference between two discrete differential operators, CDO and DDO.  They provide quantitatively similar results overall.  However, DDO performs better at interfaces damping out spurious oscillations.

The proposed methodology requires knowledge of the single crystal elastic constants. If they are not precisely known, the method could be extended to refine their values as part of the optimization procedure. 3DXRD reconstruction where grain centroids and average elastic strains/stresses are determined is now a fairly standard post-experimental analysis, and the proposed stress filtering can be directly combined with those reconstructions techniques as post-optimization correction. On the other hand, sub-grain methods are novel and still under development, and the proposed methodology could be combined with them in different ways. The sequential methodology demonstrated here, based on the application of the FFT-based micromechanical filter to stress fields resulting from diffraction-based optimization, could be extended to fulfill the diffraction and micromechanical constraints concurrently during data inversion and refinement. In both cases, but specially in the latter (concurrent optimization), the extension and application of the method has a great potential to increase both deformation and spatial accuracy, as they are interlinked, including the quantitative estimation of errors of the diffraction measurements.

In terms of numerical performance of the proposed method, since it is based on the efficient FFT computations, a single application of the filter to a $256^3$ voxel image takes 473 seconds in an Intel Core i7-8700 CPU machine. This high efficiency of a single application of the filter bodes well for standing up the more numerically involved (but still realizable) concurrent optimization, mentioned above.

As a final comment, while the proposed method is based on imposing one type of micro-mechanical constraint on the measured stress field, i.e. stress equilibrium, the dual constraint is strain compatibility. Such condition can also be expressed in terms of differential operators acting, in this case, on the total strain field. Total strains are not directly accessible by the existing diffraction-based data reduction methods, since the non-elastic part of the deformation does not directly affect lattice spacing and, consequently, diffraction patterns. However, if new methods for determining total strain fields using 3DXRD experiments are further developed, e.g. using evolving 3D microstructural images to perform DCT-based Digital Volume Correlation (DVC), in what would be an extension of similar techniques based on Computed Tomography (CT), strain compatibility imposed via FFT-based methods should be certainly considered as part of the data inversion procedures. 

\section*{Acknowledgement}  The core part of this work was conducted during HZ's visit to Los Alamos National Laboratory (LANL) to work with RAL under LANL's LDRD program, whose support is acknowledged. RAL's work is also supported by LANL's Science Campaign 2 program.  KB and HZ were also partially supported by the Air Force Office of Scientific Research through the MURI grant FA9550-16-1-0566.

\bibliographystyle{unsrt}
\bibliography{reference}

\end{document}